\pdfoutput=1
\documentclass[twocolumn,prl,amsmath,amssymb,superscriptaddress]{revtex4-1}
\usepackage[utf8]{inputenc}
\usepackage[english]{babel}
\usepackage{amsfonts,amsmath,amssymb}
\usepackage{graphicx}
\usepackage{bm} 
\usepackage{mathrsfs}
\usepackage{subfigure}
\usepackage{textcomp}
\usepackage{hyperref}
\usepackage[per-mode=symbol]{siunitx}
\usepackage[version=3]{mhchem}

\newcommand{\Iprobe}{\textup{I}_\textup{probe}}
\newcommand{\Ikick}{\textup{I}_\textup{kick}}

\DeclareSIUnit\intensity{\watt\per\centi\meter\squared}
\DeclareSIUnit\fieldstrength{\volt\per\centi\meter}

\newcommand{\degree}{\ensuremath{^\circ}}%
\newcommand{\cost}{\ensuremath{\langle\cos^2\theta_\text{2D}\rangle}}
\newlength{\figwidth}
\let\orgautoref\autoref
\providecommand{\Autoref}{%
  \def\equationautorefname{Equation}%
  \def\figureautorefname{Figure}%
  \def\subfigureautorefname{Figure}%
  \orgautoref}
\renewcommand{\autoref}{%
  \def\equationautorefname{Eq.}%
  \def\figureautorefname{Fig.}%
  \def\subfigureautorefname{Fig.}%
  \orgautoref}

\begin{document}

\title{Impulsive Laser Induced Alignment of Molecules Dissolved in Helium Nanodroplets}

\author{Dominik Pentlehner}
\thanks{These authors contributed equally}
\affiliation{Department of Chemistry, Aarhus University, 8000 Aarhus C, Denmark}

\author{Jens H. Nielsen}
\thanks{These authors contributed equally}
\affiliation{Department of Physics and Astronomy, Aarhus University, 8000 Aarhus C, Denmark}

\author{Alkwin Slenczka}
\affiliation{Institut f\"{u}r  Physikalische und Theoretische Chemie,
   Universit\"{a}t Regensburg, Regensburg, Germany}

\author{Klaus M{\o}lmer}
\affiliation{Lundbeck Foundation Theoretical Center for Quantum System
  Research, Department of Physics and Astronomy, Aarhus University, 8000 Aarhus C, Denmark}

\author{Henrik Stapelfeldt}%
\email[Corresponding author: ]{henriks@chem.au.dk}%
\affiliation{Department of Chemistry, Aarhus University, 8000 Aarhus C, Denmark}
\affiliation{Interdisciplinary Nanoscience Center (iNANO), Aarhus University, 8000 Aarhus C,
  Denmark}

\date{\today}

\begin{abstract}
  We show that a 450 fs nonresonant, moderately intense, linearly polarized laser pulse can induce field-free molecular axis
  alignment of methyliodide molecules dissolved in a helium nanodroplet. Time-resolved
  measurements reveal rotational dynamics much slower than that of isolated molecules and,
  surprisingly, complete absence of the sharp transient alignment recurrences characteristic of gas
  phase molecules. Our results presage a range of new opportunities for exploring both molecular
  dynamics in a dissipative environment and the properties of He nanodroplets.
\end{abstract}

\pacs{add pacs}

\maketitle

The ability to control how molecules are turned in space offers unique opportunities for studying
and exploiting the ubiquitous orientational dependence of a molecule's interactions with other
molecules, atoms or polarized electromagnetic radiation and for eliminating the blurring of
molecular properties and processes which usually occur in observations of randomly aligned
molecules.  For isolated molecules in the gas phase, intense laser pulses provide a versatile
approach to sharply align molecules along axes fixed in the laboratory frame
\cite{stapelfeldt_colloquium:_2003}. This ability has enabled a broad range
of new applications in molecular science
\cite{itatani_tomographic_2004,madsen_manipulating_2009,bisgaard_time-resolved_2009,wu_ultrafast_2010}.
An intriguing question is whether laser induced alignment can be extended to molecules in the
dissipative environment of a solvent. The prospects of this possibility are tremendous - notably
because most chemical reactions occur in solvents - but to date no experiments and only a few
theoretical works have addressed this issue
\cite{ohkubo_molecular_2004,ramakrishna_intense_2005,ramakrishna_dissipative_2006,hartmann_quantum_2012}.

Laser induced alignment of isolated molecules is based on a nonresonant, intense laser pulse
creating a coherent superposition of the quantum states that describe free rotation
\cite{stapelfeldt_colloquium:_2003}. Transferring the approach to molecules in a solvent faces,
however, several fundamental obstacles: \textit{(i)}, molecules in a classical solvent are not
characterized by free rotation due to collisions with the solvent. \textit{(ii)}, even if a laser
pulse could initiate coherent rotational motion the collisions would destroy it so rapidly that no
efficient alignment would be reached. \textit{(iii)}, the intense laser pulse would interact not
only with the solute molecule but also with the solvent, for instance, through polarization or even
ionization and, thereby, prevent efficient alignment. Laser induced alignment of molecules in a
solvent may thus be sensitive to and, in the worst case, hindered by numerous mechanisms which are
absent in vacuum.

The present experiment uses superfluid helium droplets
\cite{toennies_superfluid_2004,stienkemeier_spectroscopy_2006,choi_infrared_2006,braun_photodissociation_2007}
as a solvent because as a quantum fluid their unique properties strongly reduce the obstacles
\textit{(i-iii)}.  A liquid helium droplet is characterized by its ability to
solvate almost any molecule and by its low polarizability and high ionization potential, which
minimize its interaction with the laser light. The long rotational coherence times observed for
embedded molecules through spectroscopy have been interpreted as free rotation of the molecules over
several classical rotational periods within the \SI{0.37}{K} \ce{^4He} superfluid
\cite{toennies_superfluid_2004,stienkemeier_spectroscopy_2006,choi_infrared_2006}, and they build
the expectation that the He environment should be advantageous for laser induced alignment.  Our
work focuses on laser induced alignment in the nonadiabatic or impulsive regime where the laser
pulse duration is much shorter than the classical rotational period, $\text{T}_\text{rot}$, of the
molecule \cite{rosca-pruna_experimental_2001,seideman_nonadiabatic_2005}. We show that when He
droplets, singly doped with methyliodide (\ce{CH_3I}) molecules, are irradiated by a \SI{0.45}{ps}
long laser pulse the molecules reach maximum alignment \SIrange[range-phrase = --]{17}{37}{ps} later
and, in another $\sim\SI{70}{ps}$, return to random orientation which persists beyond
\SI{1000}{ps}. The dynamics differs completely from that observed for isolated \ce{CH_3I} molecules
where the first alignment maximum occurs \SIrange[range-phrase = --]{0.1}{3.5}{ps} after the center
of the pulse and recurs in regularly spaced (by $\sim\SI{33}{ps}$), narrow time windows, termed
revivals \cite{stapelfeldt_colloquium:_2003}.

\begin{figure}
  \includegraphics[width=\linewidth]{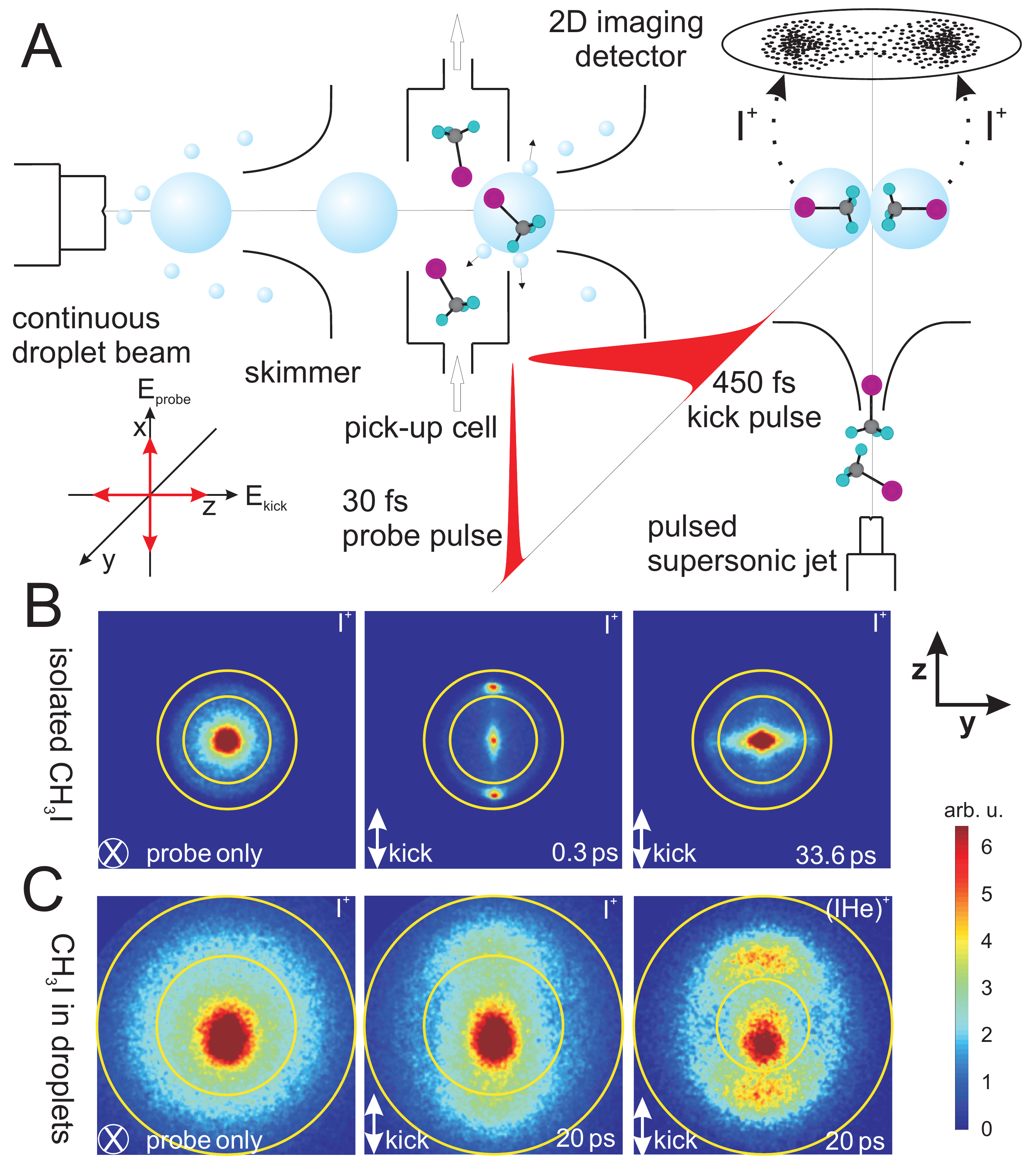}
  \caption{(A) Schematic of the key elements in the experiment. The electrostatic plates projecting
    the \ce{I^+} ions onto the imaging detector, are not shown. The polarization state of the kick
    and the probe laser pulses are shown in the coordinate system and indicated by the pulses. (B)
    \ce{I^+} ion images recorded for isolated \ce{CH_3{I}} molecules with the probe pulse only,
    $\text{t} = \SI{0.3}{ps}$ and $\text{t} = \SI{33.6}{ps}$ with the polarization of the kick pulse
    indicated. (C) \ce{I+} or \ce{IHe^+} ion images recorded for \ce{CH_3I} molecules in He droplets
    with the probe pulse only and $\text{t} = \SI{20}{ps}$. The ions detected in the area between
    the yellow circles are used to determine $\cost$. The velocity ranges depicted in all images in
    (B) and (C) are identical and the color scale for each image is normalized to the peak intensity
    outside the central spot of the image. $\Ikick =\SI{1.2e13}{\intensity}$.}
\label{fig:setup}
\end{figure}

The principle of the experiment is illustrated in \autoref{fig:setup}A. The He droplets are formed
by continuously expanding $\sim\SI{20}{atm}$ He gas, cryogenically precooled to
\SIrange[range-phrase = --]{10}{20}{K}, through a \SI{5}{\micro \meter} orifice into vacuum. The
average size of the droplets is determined by the temperature of the He gas and is $\sim \num{1.1e4}$
helium atoms per droplets at \SI{12}{K} \cite{toennies_superfluid_2004}, the temperature used for
most of our measurements. The droplet beam passes through a pick-up cell, where the partial pressure
of \ce{CH_3I} is adjusted to optimize for single molecule doping of each droplet. Hereafter, the
doped He droplet beam enters the target region where it is crossed, at \SI{90}{\degree}, by two
colinear pulsed laser beams. The first pulse (kick pulse: $\lambda = \SI{800}{nm}$, $\tau_\text{FWHM}
= \SI{450}{fs}$) is used to induce alignment. The second pulse (probe pulse: \SI{800}{nm},
\SI{30}{fs}) is used to probe the molecular orientation. It is sufficiently intense ($\Iprobe =
\SI{1.8e14}{\intensity}$) to multiple ionize \ce{CH_3I} resulting in Coulomb explosion into
positively charged fragments, notably \ce{I^+} with a \ce{CH_3^+} partner. These \ce{I^+} ions
recoil along the molecular symmetry axis of the \ce{CH_3I} parent molecule and by detecting their
emission direction with a velocity map ion imaging spectrometer we determine the spatial orientation
of \ce{CH_3I} at the time defined by the delay, t, between the kick and the probe pulse
\cite{larsen_aligning_1999}. An important part of the experiment is the ability to compare alignment
of molecules embedded in the droplets with alignment of isolated molecules under identical
conditions of the laser pulses. Therefore, a second molecular beam, containing isolated \ce{CH_3I}
molecules, can be sent into the target region (see \autoref{fig:setup}).


\Autoref{fig:setup}B shows examples of \ce{I^+} ion images from Coulomb explosion of isolated
\ce{CH_3I} molecules. When only the probe pulse is used the \ce{I^+} image is circularly symmetric
showing that the molecules are randomly oriented as expected (left). When the kick pulse is
included the \ce{I^+} ions localize along (middle) or perpendicular (right) to the kick
pulse polarization at certain times. At these times the symmetry axis of the molecule is aligned or
antialigned, respectively. The degree of alignment is quantified by determining the average value of
$\theta_\text{2D}$, $\cost$, where $\theta_\text{2D}$ is the angle between the kick pulse
polarization and the projection of an \ce{I^+} ion recoil velocity vector on the detector screen. It
is determined only from those ions pertaining to the directional \ce{I^+} -- \ce{CH_3^+} Coulomb
explosion channel marked by circles in \autoref{fig:setup}B.

\begin{figure}
  \includegraphics[width=\linewidth]{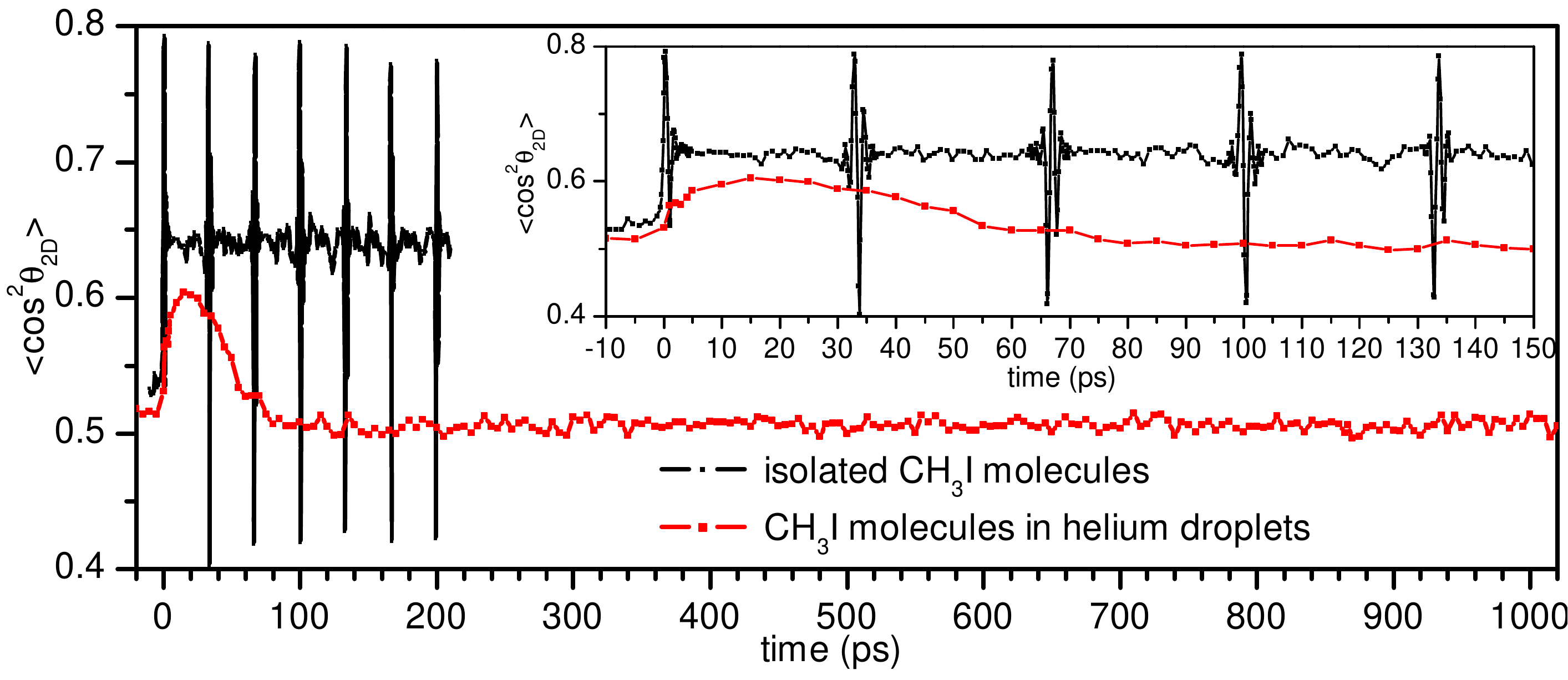}
  \caption{The degree of alignment, represented by $\cost$ determined from \ce{I+} images, of isolated
    \ce{CH_3I} molecules (black squares) and of \ce{CH_3I} molecules in He droplets (red squares) as
    a function of time after the peak of the kick pulse (centered at t = 0). The parameters of the
    kick and the probe pulse are identical for the two data series. $\Ikick =
    \SI{1.2e13}{\intensity}$. The inset expands on the first \SI{150}{ps}.}
  \label{fig:align_long}
\end{figure}

The time dependence of $\cost$ for isolated molecules is displayed in \autoref{fig:align_long}
(black curves) and it can be explained by the formation of a rotational wave packet by the kick
pulse \cite{seideman_nonadiabatic_2005}.  When the kick pulse is applied $\cost$ rises sharply to a
local maximum at $\textup{t} = \SI{0.3}{ps}$.  Hereafter, it drops quickly to a value of 0.64 and
remains at this level except in narrow time windows, separated by $\sim\SI{33.3}{ps}$. These
transients are the manifestations of full revivals $(\text{t} = \SI{0.3}{ps} + \text{N}
\text{T}_\text{rot},~\text{N}=1,2,\ldots)$ and half-revivals $(\text{t} = \SI{0.3}{ps} +
(\text{N}-{1}/{2})\text{T}_\text{rot},~\text{N}=1,2,\ldots)$ where $\text{T}_\text{rot} = 1 /
{\text{2Bc}} = \SI{66.7}{ps}$ for a rotational constant $\textup{B} = \SI{0.250}{cm^{-1}}$.  The
observations are fully consistent with many previous studies of impulsive alignment
\cite{rosca-pruna_experimental_2001,hamilton_alignment_2005}.

Turning to the doped He droplet data we note that the I$^+$ image recorded with the
probe pulse only (\autoref{fig:setup}C, left) is also circularly symmetric. The radial distribution,
i.e. the momentum distribution, is, however, much broader.
After the kick pulse, the \ce{I+} ions localize along its polarization axis (middle panel), which is a clear
manifestation of alignment. The same information is obtained from \ce{(IHe_n)^+} images, which only
have a narrower radial distribution (right panel). The slight off-set of the ion distributions from
the center of the images in panel C results from the velocity of the droplet beam (along the z-axis)
which, unlike the beam of isolated molecules, is parallel to the ion detector -- see
\autoref{fig:setup}A.

The time dependence of $\cost$ for \ce{CH_3I} molecules in He droplets is displayed by the red
curves in \autoref{fig:align_long}. After $\sim \SI{20}{ps}$ $\cost$ reaches a maximum of 0.60
demonstrating unambiguously that molecules can be aligned inside a He droplet under field-free
conditions. The $\cost$ curve for the molecules in the droplets differs strongly from the curve
recorded for the isolated molecules. Not only is the rise time to the first alignment peak much
longer ($\sim\SI{20}{ps}$ versus \SI{0.3}{ps}), it also decreases to a value of 0.5 with an
isotropic image, which characterizes a sample of randomly oriented molecules, within
$\sim\SI{100}{ps}$. For the isolated molecules the prompt alignment decays within $\sim \SI{4}{ps}$
to a $\cost$ value of 0.64, characterizing the time-independent permanent
alignment~\cite{seideman_nonadiabatic_2005}, and regains high values at the revivals. For molecules
in the droplets $\cost$ stays constant at 0.5 for $\textup{t} = \SI{100}{ps}$ to $\SI{1000}{ps}$
without any significant signs of revivals. Finally, the maximum degree of alignment observed, 0.60,
falls below the maximum value of 0.78 obtained for the isolated molecules and stays slightly below
the 0.64 of the permanent alignment.

Our observations may at first seem at odds with expectations based on the prevailing conception of
rotational structure and dynamics of molecules inside He droplets. Indeed high resolution infrared
and microwave absorption spectra have established that a gas phase Hamiltonian describes the
rotational spectrum well with the effect of the He solvent being to reduce (increase) the rotational
constants (moments of inertia) by up to a factor of $\sim 5$. This is commonly attributed to
adiabatic following of a nonsuperfluid He solvation shell surrounding the solute
molecule~\cite{choi_infrared_2006,lehnig_rotational_2009}. Thus, one might expect that
the alignment dynamics would be similar to that observed for isolated
molecules but slowed down by up to a factor of $\sim 5$ due to the increased moments of
inertia. Clearly, this is not what is observed.

\begin{figure}
  \includegraphics[width=\linewidth]{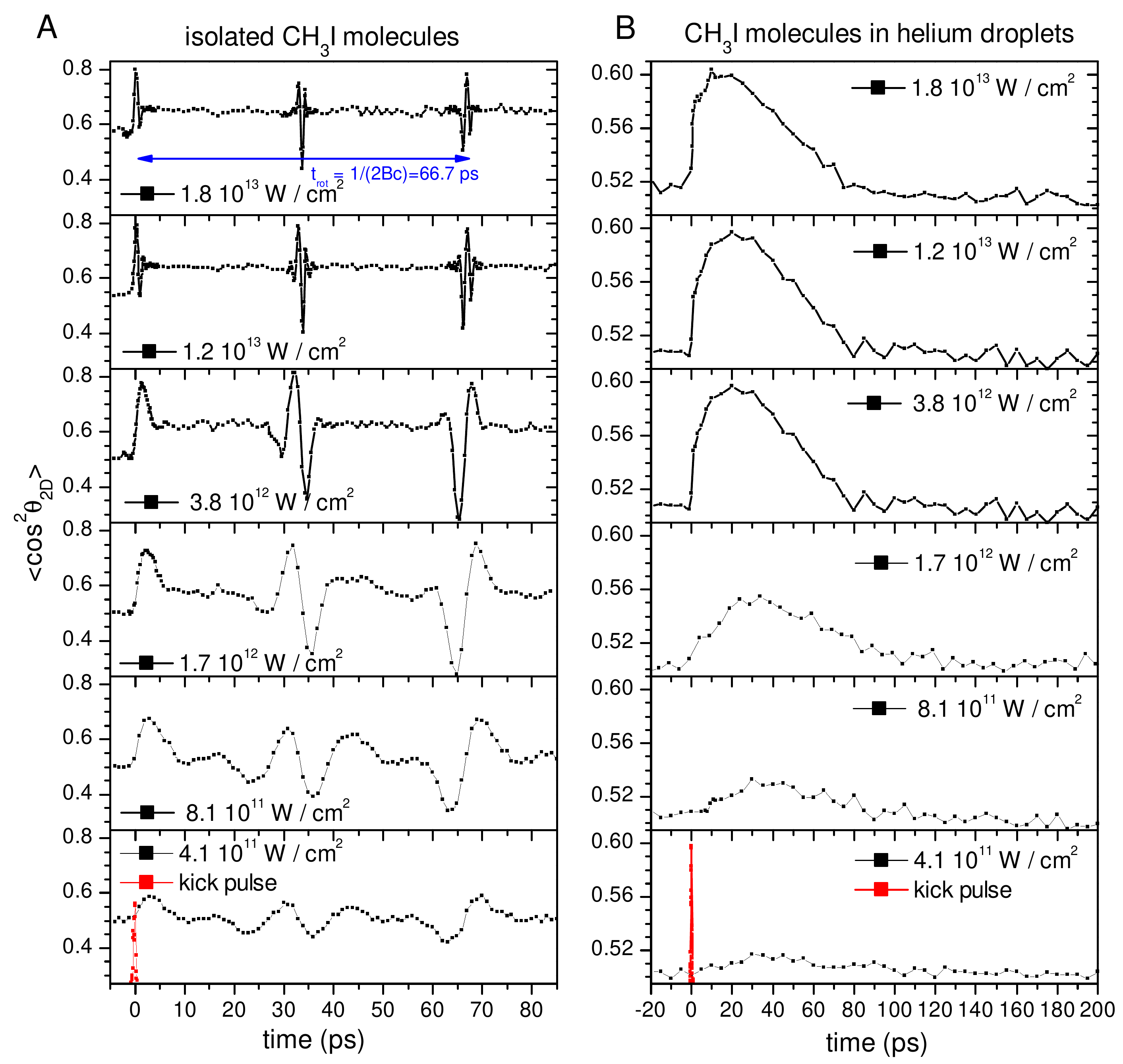}
  \caption{$\cost$ as a function of time, recorded at different intensities of the kick pulse and
    for both isolated \ce{CH_3I} molecules (A) and \ce{CH_3I}
    molecules in He droplets (B).}
  \label{fig:allinone}
\end{figure}

To gain further insight into laser induced molecular alignment in He droplets the time dependence of
$\cost$ was recorded at different intensities of the kick pulse. In particular, the measurements
were pushed towards as low intensities as possible aiming at forming a wave packet composed of low
lying rotational states, since these states have the longest lifetimes according to spectroscopic
studies~\cite{choi_infrared_2006}. Thus, coherence between the rotational states should be allowed
to survive sufficiently long that revivals could be formed. The reference measurements on isolated
molecules (\autoref{fig:allinone}A) show that the revivals evolve from very sharp to very broad
structures as the intensity is lowered which is a clear manifestation that the wave packet changes
from comprising a broad superposition of high lying states to just a few low lying states. In
addition, the permanent alignment level, observed between the revivals, decreases due to excitation
into gradually lower rotational states. The data for \ce{CH_3I} in He droplets
(\autoref{fig:allinone}B) show that the maximum degree of alignment gradually occurs at later times
and gradually decreases as the intensity is lowered.  There is, however, no sign of revivals at any
of the intensities and the peak value stays slightly lower than the permanent alignment in the
corresponding isolated molecule case.

\begin{figure}
  \includegraphics[width=0.9\figwidth]{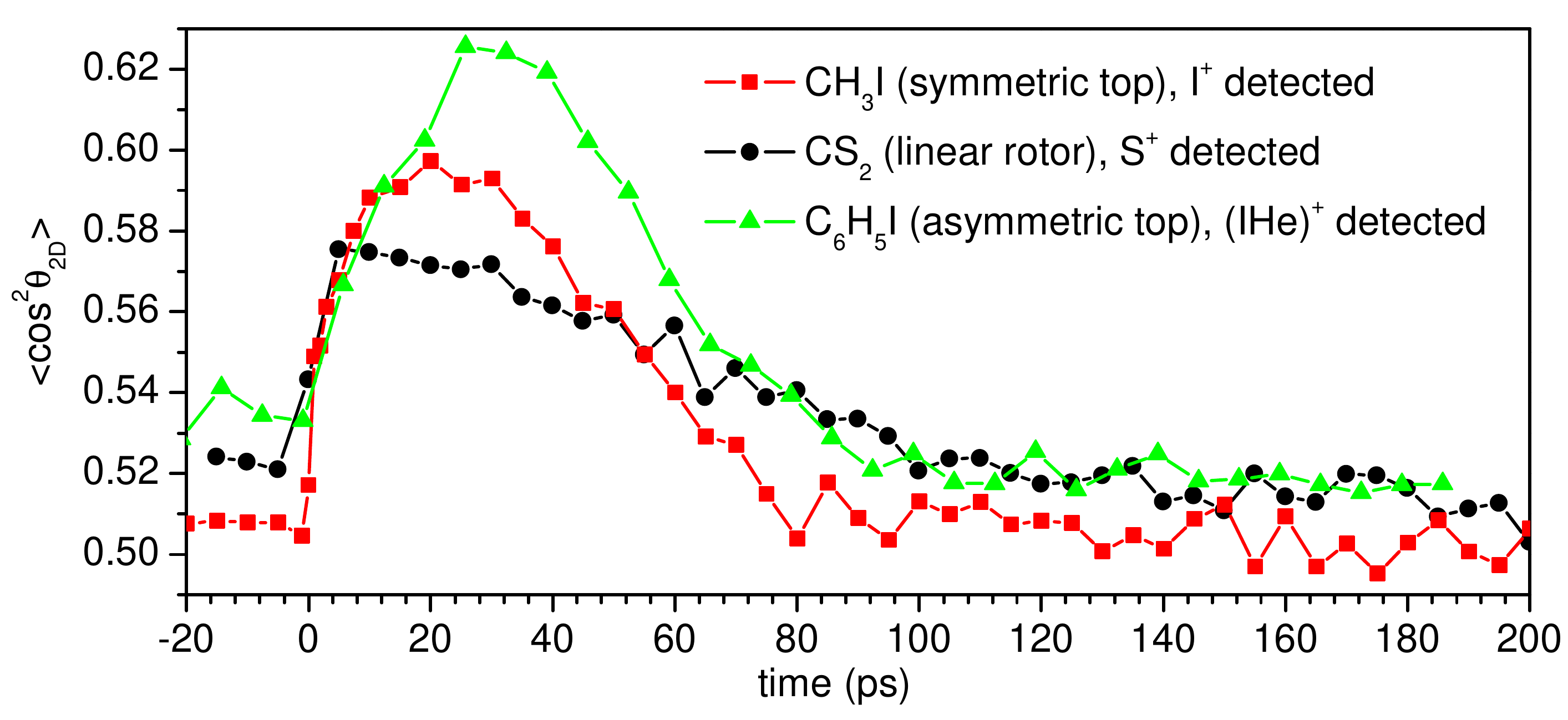}
  \caption{$\cost$ as a function of time, recorded for \ce{CH_3I} (red squares), \ce{CS_2} (black
    circles) and \ce{C_6H_5I} (green triangles) in He droplets. $\Ikick = \SI{1.2e13}{\intensity}$.}
  \label{fig:align_molecules}
\end{figure}

To ensure that the alignment dynamics observed is not a result of some unusual behavior of
\ce{CH_3I} in He droplets, we repeated the measurements for carbondisulfide (\ce{CS_2}) and
iodobenzene (\ce{C_6H_5I}) molecules.  The time dependence of $\cost$ for all three molecules is shown
in \autoref{fig:align_molecules}. Although there are minor differences between the three curves, the
overall alignment dynamics is very similar, whereas their alignment dynamics in the gas phase is
significantly
different~\cite{hamilton_alignment_2005,kumarappan_aligning_2007,holmegaard_control_2007}. Furthermore,
we explored if the alignment dynamics observed is sensitive to the size of the He droplet. In
practice this was done by recording the time dependence of $\cost$ for \ce{CH_3I} in He droplets at
temperatures of 12, 16 and \SI{18}{K} of the expanding $\sim\SI{20}{atm}$ He gas. This corresponds
to mean droplet sizes of $\sim 11000$, $\sim 2000$, and $\sim 900$ He atoms per droplet,
respectively~\cite{toennies_superfluid_2004}. No significant differences between the three curves
could be detected.  The absence of any effect of the droplet size rules out several plausible
explanations for the observed dynamics such as an inhomogeneous effect due to the distribution of
the droplet size, an influence of translational states of the embedded molecule within the droplet,
a coupling to ripplons \cite{stienkemeier_spectroscopy_2006}, or an overall rotation or deformation
of the whole droplet.

The observed effects are not captured by the published theoretical works on the influence of a
dissipative environment on nonadiabatic alignment of
molecules~\cite{ramakrishna_intense_2005,ramakrishna_dissipative_2006,hartmann_quantum_2012}. In
these works, correlations between the molecule and the surrounding particles are neglected, and the
environment is taken into account by incorporating decoherence and population decay of the molecular
rotational wave packet -- induced by elastic and inelastic collisions -- only \emph{after} the pulse
is switched off.  The rotational wave packet coherence can indeed be lost due to such collisions
\cite{vieillard_field-free_2008,owschimikow_cross_2010}, but they would affect only the amplitude
and not the period of the rotational motion
\cite{ramakrishna_dissipative_2006,hartmann_quantum_2012,owschimikow_cross_2010,blokhin_intermolecular_1999}. This
contradicts the observed dynamics which is slower than the anticipated wave packet dynamics even
when the moment of inertia is multiplied by a factor of 5 compared to the gas phase value. Thus, the
comparison of our results with the theoretical works indicate that interactions \emph{during} the
pulse and, possibly, correlations between the molecule and the helium droplet are influencing the
dynamics.

This hypothesis is corroborated by comparing our experiment with microwave-
\cite{lehnig_rotational_2009} and infrared-spectroscopy \cite{choi_infrared_2006} of molecules in
helium droplets: At the lowest kick pulse intensities, only the lowest rotational states ($\text{J} \le 4$),
with long coherence lifetimes~\cite{choi_infrared_2006}, are excited. The intensity of the kick
pulse ($\sim\SI{e11}{\intensity}$) is, however, orders of magnitudes higher than the intensities
($\lesssim\SI{e2}{\intensity}$) used in spectroscopy. Rather than spectroscopically probing the
rotational eigenstates of the coupled system, the kick pulse impulsively excites rotational motion
of the molecule within the droplet, and the subsequent dynamics is not consistent with the long
coherence times observed in spectroscopy. At the higher kick pulse intensities, we enter a regime of
rotational states that is not investigated in He droplet spectroscopy and whose energies and
lifetimes are not known. The absence of any discontinuity in the dependence on the kick pulse
intensity (\autoref{fig:allinone}B) indicates, though, that similar mechanisms may explain the
experiments at all intensities, and thus a complete understanding of the highly excited J-states may
not be needed to interpret our results.

Our molecules are subject to a strong laser field, an impulsive excitation of the molecular motion,
and the interaction with the helium environment. Comparison to impulsive alignment of isolated
molecules and to spectroscopy in droplets suggests that the reason for the unexpected rotational
dynamics lies in the combination of the three. In particular, the time-dependence of the alignment
at different kick pulse intensities and the only marginal dependence on the molecular species
indicate that the droplet response to the molecular excitation is significantly influencing the
observed dynamics. One potential effect is the exchange of angular momentum between the molecule and
the helium environment -- an effect starting already during, and possibly enhanced by, the kick
pulse. The resulting scrambling of M values (M being the projection of the rotational angular
momentum on the axis defined by the kick pulse polarization) implies that the rotational wave packet
formed differs from the one created in isolated molecules (where $\Delta \text{M} =0$
\cite{hamilton_alignment_2005}) and may partly explain the slow rise and the weaker degree of the
alignment. The M scrambling will continue after the kick pulse and together with the onset of
population decay of rotational states it ultimately forces the alignment to decay towards the value
of an isotropic ensemble - as observed. In addition, the impulsive excitation of the molecule may
excite collective modes of the helium droplet which will be correlated with the different components
of the rotational wave packet and, thereby, also influence the alignment dynamics. A full account of
the details of the observed dynamics requires a quantitative many-body analysis of the coherent and
dissipative molecule-solvent interactions, which is beyond the scope of this letter.

To conclude, we have shown that a short, intense laser pulse can align molecules dissolved in He
nanodroplets under field-free conditions. This opens many new opportunities, notably for
time-resolved imaging of molecular frame reaction dynamics under influence from the dissipative
environment of a solvent \cite{gruner_vibrational_2011}. The ability of He droplets to dissolve
molecules ranging from the smallest diatomics over biochromophores
\cite{smolarek_high-resolution_2010} to proteins \cite{bierau_catching_2010} presage studies for a
variety of species. The understanding of the slow rotational dynamics and absence of revivals
reported here calls for new theoretical descriptions, which could provide novel insight into the
elementary excitations of He nanodroplets. In particular, rotational wave packets can transfer large amounts of angular momentum to the He surroundings
and may be a way to explore quantized vortices
\cite{stienkemeier_spectroscopy_2006,dalfovo_pinning_2000,madison_vortex_2000,gomez_traces_2012},
damping of collective modes, and their relation to the critical velocity in superfluid He droplets
\cite{burger_superfluid_2001}.

The work was supported by the Danish Council for Independent Research (Natural
Sciences), The Lundbeck Foundation and the Carlsberg Foundation.

\end{document}